\documentclass[%
reprint,
 amsmath,amssymb,
 aps,
 pra, 
floatfix,
]{revtex4-1}

\usepackage{color,soul}
\usepackage{graphicx}
\usepackage{dcolumn}
\usepackage{bm}

\begin{document}


\title{Activity-assisted self-assembly of colloidal particles}

\author{S.A. Mallory}
\author{A. Cacciuto}%
 \email{ac2822@columbia.edu}
\affiliation{%
Department of Chemistry, Columbia University\\ 
3000 Broadway, New York, NY 10027
}%





\date{\today}

\begin{abstract}
We outline a basic strategy of how self-propulsion can be used to improve the yield of a typical colloidal self-assembly process. The success of this approach is predicated on the thoughtful design of the colloidal building block as well as how self-propulsion is endowed to the particle.  As long as a set of criteria are satisfied, it is possible to significantly increase the rate of self-assembly, and greatly expand the window in parameter space where self-assembly can occur.  In addition, we show that by tuning the relative on/off time of the self-propelling force it is possible to modulate the effective speed of the colloids allowing for further optimization of the self-assembly process. 

\end{abstract}

\maketitle


The colloidal self-assembly of a well defined micro-structure is a complex process that is difficult to control with any degree of precision.  Most modern approaches to this problem revolve around tailoring the morphology of the individual colloids.  By either altering the shape of the colloid or introducing highly selective interactions, it is possible to drive the mono-disperse self-assembly of a specific colloidal cluster.  This approach has been particularly fruitful and some of the more prominent examples include: patchy particles \cite{zhang1407,zhang11547,wilber085106,glotzer419,pawar150,romano171,yi193101,glotzer557,chen381}, lock and key colloids \cite{sacanna575,bowden233,sacanna1688,ashton9661,sacanna96,sacanna8096,wang51}, and DNA decorated colloids \cite{biancaniello058302,valignat4225,nykypanchuk549}. Nevertheless, it is well known that high yield self-assembly only occurs for very specific particles shapes and interaction strengths, making the search for the self-assembly window in parameter space rather cumbersome. This is typically due to the presence of large kinetic and entropic barriers, and in many cases it is only possible to obtain a meaningful yield of the target structure after waiting a significant amount of time.   

Recent advances in colloidal chemistry have led to the reliable synthesis of self-propelled or active colloids \cite{theurkauff268303,paxton6462,ginot011004,palacci936,palacci20130372,cohen068302,jiang268302,buttinoni284129,brown4016,valadares565,gao467,Wang17744}. These particles can be thought of as the colloidal analog of swimming bacteria and are characterized by inherently non-equilibrium, directional forces that can propel them at velocities of tens of microns per second. Recent experimental and numerical studies have considered a myriad of objects immersed in these suspensions of self-propelled colloids, which are often referred to as active fluid \cite{harder062312,harder194901,kaiser1275,mallory012314,mallory032309,kaiser260,ni018302,li224903,shin113008,kaiser124905,Angelani048104,DiLeonardo9541,Leptos198103,Dunkel4268,Morozov2748,ray013019,molina1389,angelani138302,mino1469,kasyap081901,kaiser158101,wu3017,mino048102,eckhardt96,koumakis,krafnick022308,jepson041002,kaiser044903,angelani113017,takagi1784,farage042310}.  It has been shown by multiple investigators that active fluids exhibit unique thermo-mechanical properties \cite{mallory052303,solon198301,ginot011004,solon673,smallenburg032304,ezhilanR4,yanR1,joyeux032605,yang6477,fily5609,winkler6680,bertin44,takatori028103, wykes4584} and are a powerful medium for mediating the effective interactions between suspended passive colloids and polymers. In addition, active fluids have proven to be a useful tool for powering primitive micro-machines and controlling the transport properties of passive tracers.
\begin{figure}[b]
\includegraphics[width = 3.0in]{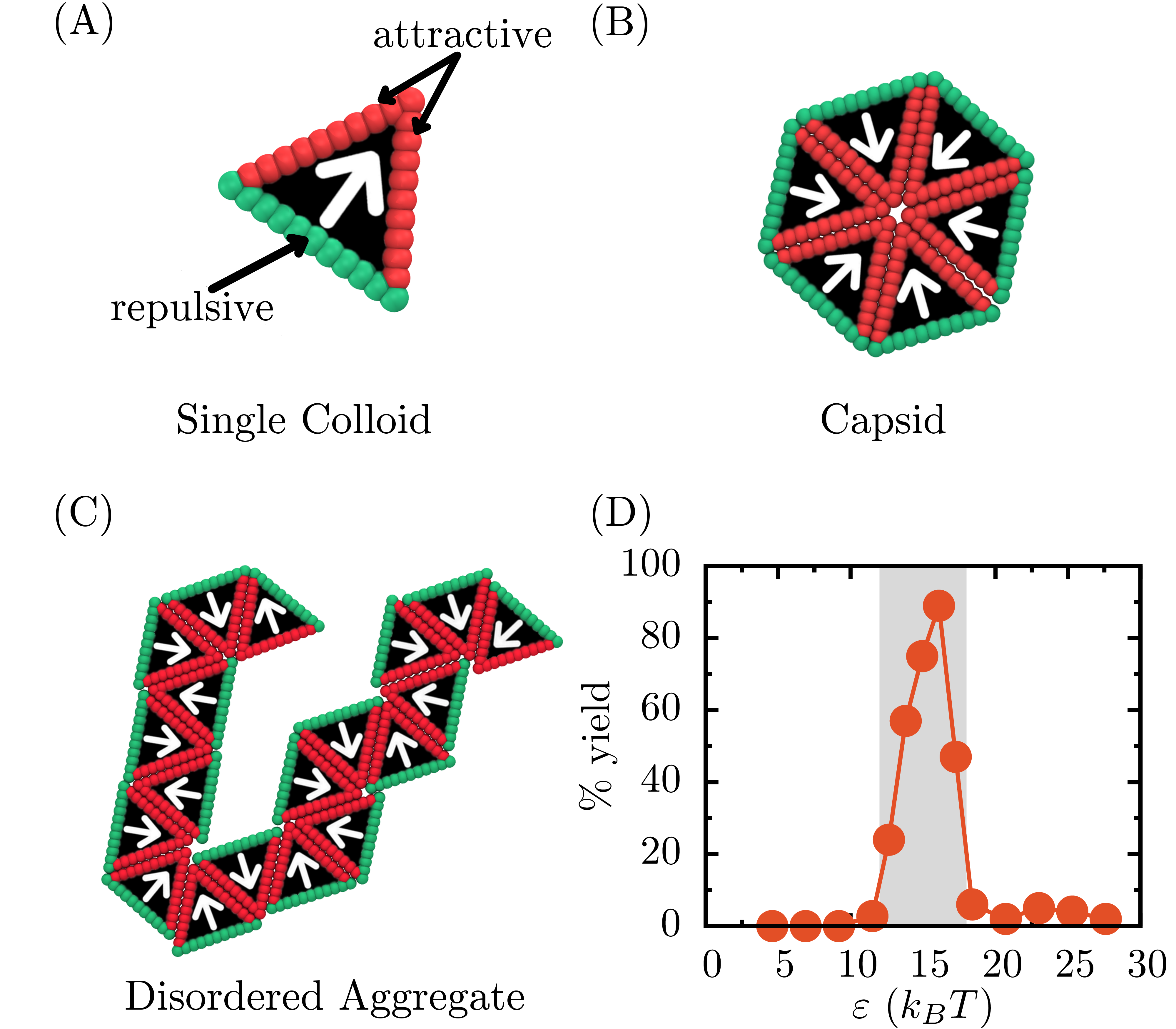}

\caption{(A)   The colloidal building blocks have been functionalize such that two of the three faces exhibit a short range attraction.  The white arrow indicates the direction of self-propulsion. (B)  The target structure to be self-assembled is a hexagonal capsid composed of six colloidal building blocks. (C) Typical disordered aggregate formed when the attraction between colloids is too strong.  (D) Percent yield of target structure as a function of the strength of the attractive interaction $\varepsilon$ for passive colloids $(F_a \sigma = 0\  k_BT)$. }

\label{fig1}

\end{figure}
In this work, we explore how self-propulsion can be used as an extra handle for colloidal self-assembly. By introducing a new dimension in the parameter space, we show how it is possible to bias the selection of a particular micro-structure and greatly improve the overall yield and  rate of the assembly process.  The main idea behind this approach is to use the self-propelling forces generated by the colloids to reinforce the stability of the desired micro-structure while simultaneously destroying any malformed or competing micro-structures.  This technique not only expedites the rate of self-assembly, but broadens the window in parameter space where self-assembly is actually possible.  

 As our colloidal building blocks we functionalize equilateral triangles with the aim of self-assembling capsid like structures. The design of the colloidal building block and the target structure are shown in Figs. \ref{fig1}A and \ref{fig1}B, respectively. The formation of the capsid is driven by both the shape of the colloid and an anisotropic interaction between colloids.  Each colloid is patterned such that two of the three faces exhibit a short range attraction. There are three reasons for choosing this particular shape and surface interaction. The first being that there is a clearly defined target structure.  Given the tailored nature of the interaction between colloids, one can readily expect them to self-assemble into finite-sized closed hexagonal aggregates. The second reason  is that there exist an ensemble of structures with a larger degree of orientational entropy (see Fig.~\ref{fig1}C) that can directly compete with the formation of the target structure.  This high level of competition generates a scenario where the success of self-assembly is highly dependent on the strength of attraction between colloids and is expected to only occur for a narrow range of parameters.  Lastly, we have chosen these specific building blocks because the vectorial sum of the colloids' propelling axes in the target structure is equal to zero (see Fig.~\ref{fig1}B), and they all point towards the center of the aggregate. This will turn out to be a crucial requirement if self-propulsion is to improve the self-assembly process. 

The model and simulation technique implemented here are similar to the the work of Zhang et al.~\cite{zhang1407} where they studied the self-assembly behavior of a similarly shaped patchy particle.  The clustering behavior of a similar shaped active particle was also recently investigated \cite{ilse2016}.  A complete description of the simulation details is included in Appendix A.  To summarize, each colloid is confined to move in two dimensions, has mass $m$, and undergoes translational and rotational Langevin dynamics at a constant temperature $T$. Self-propulsion is introduced through a directional force of constant magnitude $F_a$ and is directed along a predefined orientation vector $\boldsymbol{n}$ which passes through the center of mass of each particle and is perpendicular to the purely repulsive face of the colloid as illustrated in Fig.~\ref{fig1}A. The edge length of the colloid is fixed at $2 \sigma$, where $\sigma$ is the unit length in our simulations. Here, we define the  P\'eclet number of an individual colloid as ${\rm Pe}=v_0 \frac{\tau_D}{\sigma}= \frac{F_a\sigma}{k_BT}$, where  $\tau_{D}=\sigma^2/D$ is the particle self-diffusion time, $D=\frac{k_BT}{\gamma}$ is the linear translational diffusion constant,and $v_0$ is the swim velocity of the colloid, which is related to the propelling force via $v_0=F_a/\gamma$ where $\gamma$ is the friction coefficient. The attraction between the faces of the colloid is quantified in terms of the binding energy when a pair of colloids have their attractive faces fully aligned and in contact which we denote by $\varepsilon$. The attraction is quite short ranged and decays to zero within a distance $0.5\sigma$ from the face of the colloid.  (See supplement for full details). Using the numerical package LAMMPS ~\cite{plimpton1}, all simulations were carried out in a periodic box of dimension $L$ with $T=m=\sigma=\tau=1$. All quantities in this investigation are given in reduced Lennard-Jones units. 

We begin our analysis by identifying the range of binding energies $\varepsilon$ for which passive colloids successfully self-assemble into the target structure.  A series of numerical simulations for different values of $\varepsilon$ were carried out in the absence of any self-propelling force $(F_a \sigma = 0\  k_BT)$. In each simulation, the total number of colloids is fixed at $N=600$ with an overall volume fraction of $\phi = 0.1$.  We restrict our study to this dilute regime as it predisposes hexagonal assembly over the formation of disordered aggregates.  Each simulation is run for a minimum of $10^8\tau$. In the case of perfect self-assembly, we would expect to form $100$ capsid structures.  We define the yield of the process as the average number of successfully self-assembled structures out of the theoretical maximum number of target structures.  Figure~\ref{fig1}D shows the percent yield of hexagonal structures as a function of the strength of the attractive interaction. As expected, only a narrow range of  $\varepsilon$ returns a respectable yield, with a maximum of 90\% at around $\varepsilon \approx 16 \   k_BT$. For values of $\varepsilon$ smaller than this range, the colloids do no aggregate and the system remains in the fluid phase. For larger values of $\varepsilon$, the formation of large disordered aggregates is favored (see Figure \ref{fig1}C).
\begin{figure}[b]
\includegraphics[width = 3.0in]{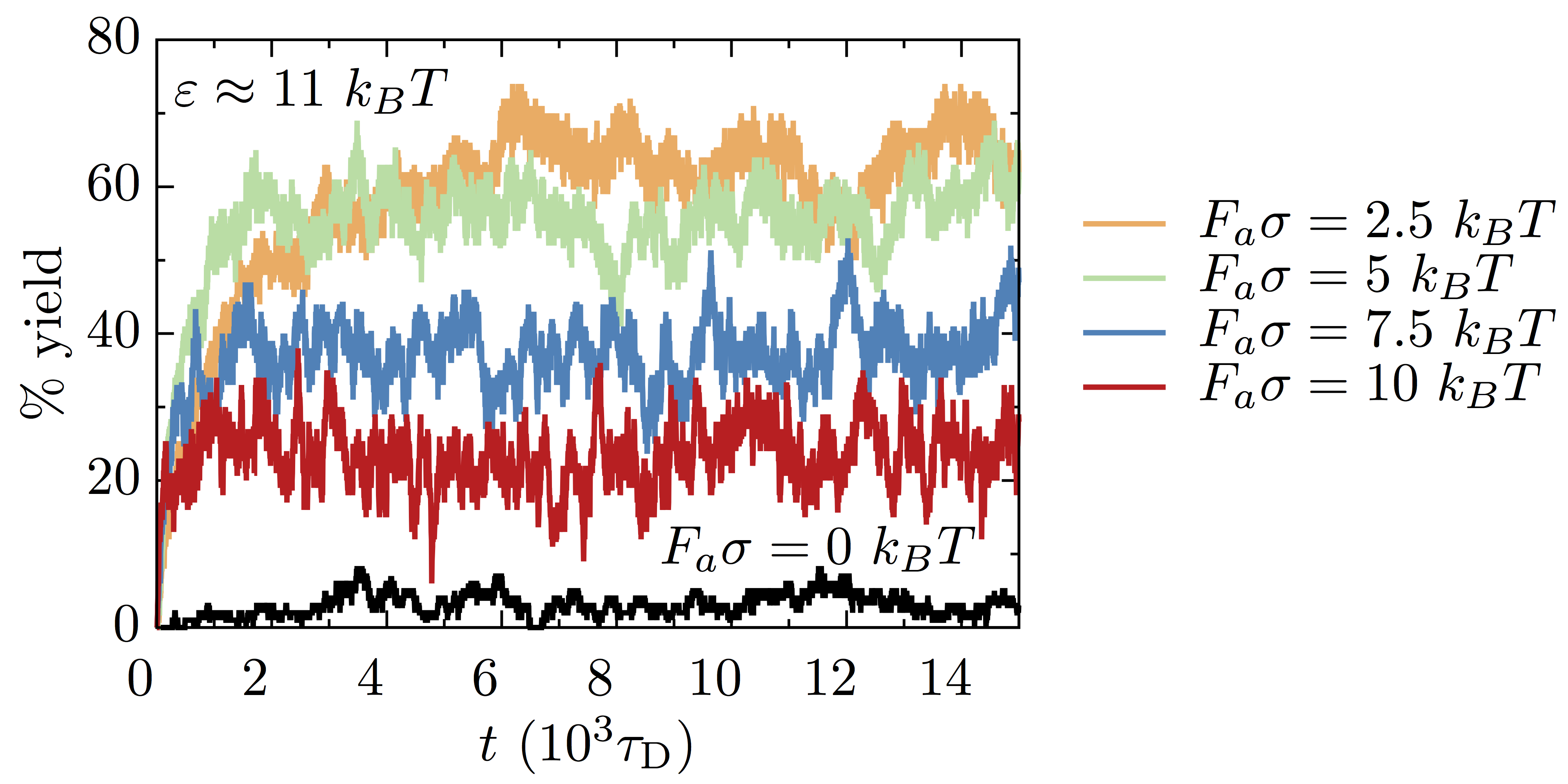}
\caption{Percent yield of the target structure as a function of time for binding energy $\varepsilon \approx 11 \  k_BT$ and several values of the self-propelling force $F_a$.}

\label{fig2}

\end{figure}

A similar set of simulations were also carried out where a self-propelling force $F_a$ was applied to each colloid.  In Figure \ref{fig2}, we plot typical trajectories of the percent yield as a function of time for several values of the self-propelling force $F_a$. To illustrate the beneficial effects self-propulsion can have on self-assembly, we consider the case where the attraction between colloids ($\varepsilon \approx 11 \  k_BT$) is too weak to drive a significant aggregation in the absence of active forces.  In this instance, the yield never rises above about $5\%$.  By introducing a small amount of self-propulsion to the colloids ($F_a \sigma = 2.5 \  k_BT$ and $F_a \sigma = 5.0 \  k_BT$), it is possible to dramatically improve self-assembly.  For both of these values the yield increases at a rapid rate before plateauing at around $60\%$. The yield does however begin to rapidly decreases if the activity of the colloids is increased any further (i.e. $F_a \sigma > 7.5 \ k_BT$).  For these larger values of $F_a$, the forces exerted by a single active colloid swimming in the bulk are able to destroy a fully formed target structure.  The rate of formation of the target structure is self-regulated by these errant colloids eventually leading to a steady state condition where the percentage of target structures formed fluctuates about some fixed value.  This study suggests that a) it is possible to significantly improve the self-assembly of these colloids by introducing a judicious amount of self-propulsion even in regions of parameter space where passive colloids wouldn't spontaneously self-assemble,  b) the overall dependence of the yield on the self-propelling force is non-monotonic (i.e. there is a range of self-propelling forces for which the maximum yield can be obtained, but above which a strong decline is  expected).
\begin{figure}[b]
\includegraphics[width = 3.0in]{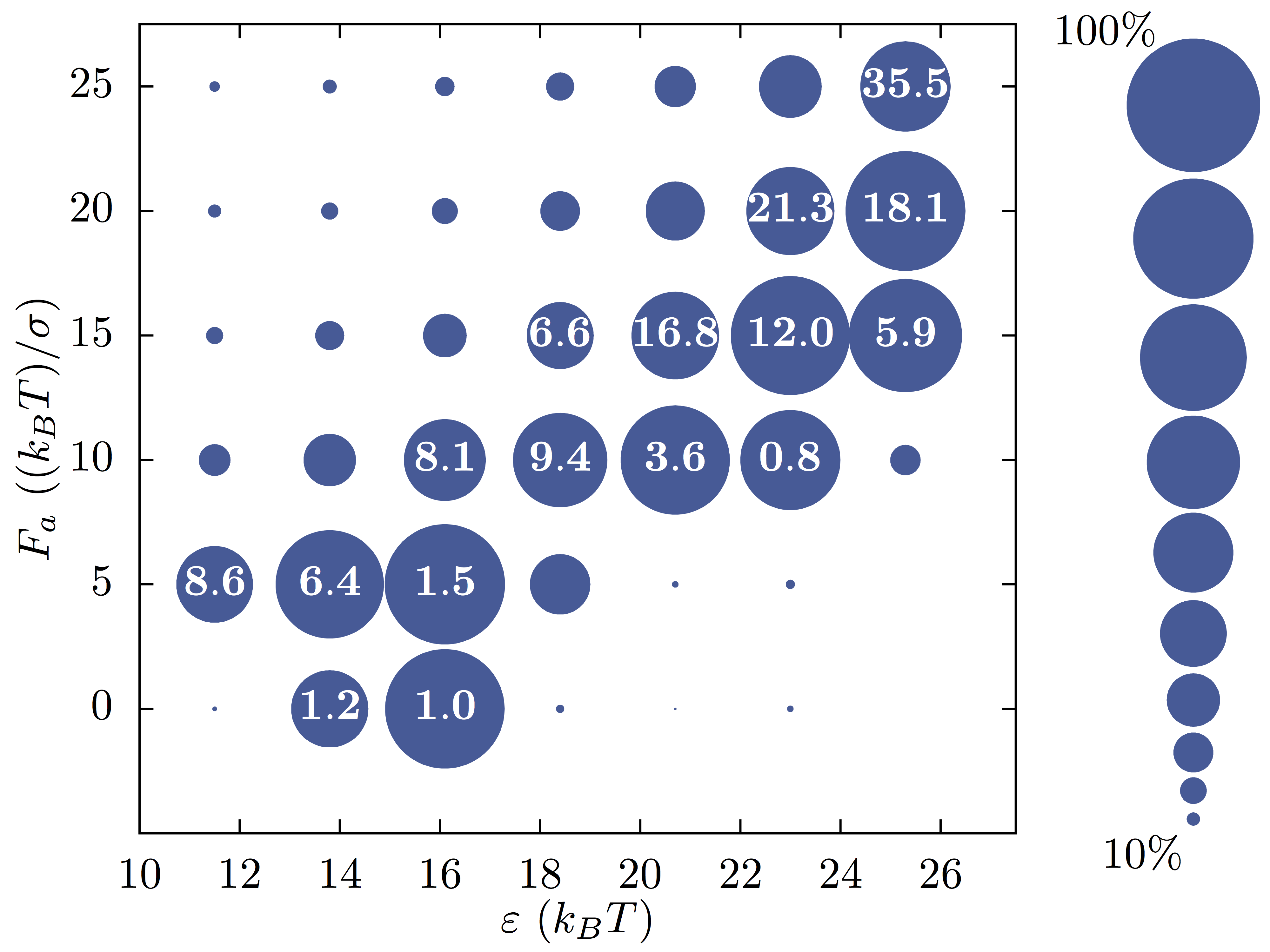}
\caption{Percent yield of the target structure as a function of the binding energy $\varepsilon$ and the self-propelling force $F_a$. The white number at the center of the circle indicate the relative increase in the rate of the assembly process, $\nu_{\rm SA}$, with respect to the best performing passive case that occurs for $\varepsilon \approx 16 \  k_BT$.}
\label{fig3}
\end{figure}

Figure \ref{fig3} summarizes all our results into a two dimensional plot where we report the equilibrium yield as a function of the binding energy $\varepsilon$ and the self-propelling force $F_a$.  Crucially, the size of the region in parameter space where self-assembly takes place is now much wider for self-propelled systems.  In many instances, the self-propelled systems can achieve yields comparable or greater than those obtained for the optimal passive case.  We find that respectable yields of the target structure are typically obtained when the self-propelling force is slightly weaker than the binding energy between two colloids (i.e. $(F_a\sigma)/\varepsilon \lesssim 1$).  As a measure of the rate of the self-assembly process, we measure $\tau_{\frac{1}{2}}$, defined as the time required for the yield to reach $50\%$ of the possible maximum yield, and then compute the relative self-assembly rate $\nu_{\rm SA}=\tau^*_{\frac{1}{2}}/\tau_{\frac{1}{2}}$, where $\tau^*_{\frac{1}{2}}$ corresponds to the $\tau_{\frac{1}{2}}$ obtained for the best performing passive system ($\varepsilon \approx 16 \  k_BT$).  Remarkably, it is possible to significantly increase the rate of self-assembly without compromising the yield by using the largest possible values of $F_a$ and $\varepsilon$  for which  $(F_a\sigma)/\varepsilon \lesssim 1$.  In other words, fast and sticky colloids give the overall best self-assembly results in terms of speed and yield (at least within the range of parameters considered in this work).
 
The underlying mechanism responsible for improving self-assembly can be understood with simple geometric arguments.  As mentioned above, a critical requirement of the target micro-structure is that the vectorial sum of the self-propelling forces pointing in its interior is equal to zero.  This creates a focal point in the center of the compact aggregate where each colloid can exert a force $F_a$ that strengthen their mutual attractive interactions.  Also, any aggregate for which this vectorial condition is not satisfied, will experience large active torques and shear forces that can break them apart.  One can think of self-propulsion in these systems as a very selective filter that only allows for the stabilization of certain structures.  We believe that this is a general feature of this approach and should work for all compact target structures satisfying the vectorial condition discussed above.

\begin{figure}[t]
\includegraphics[width = 2.5in]{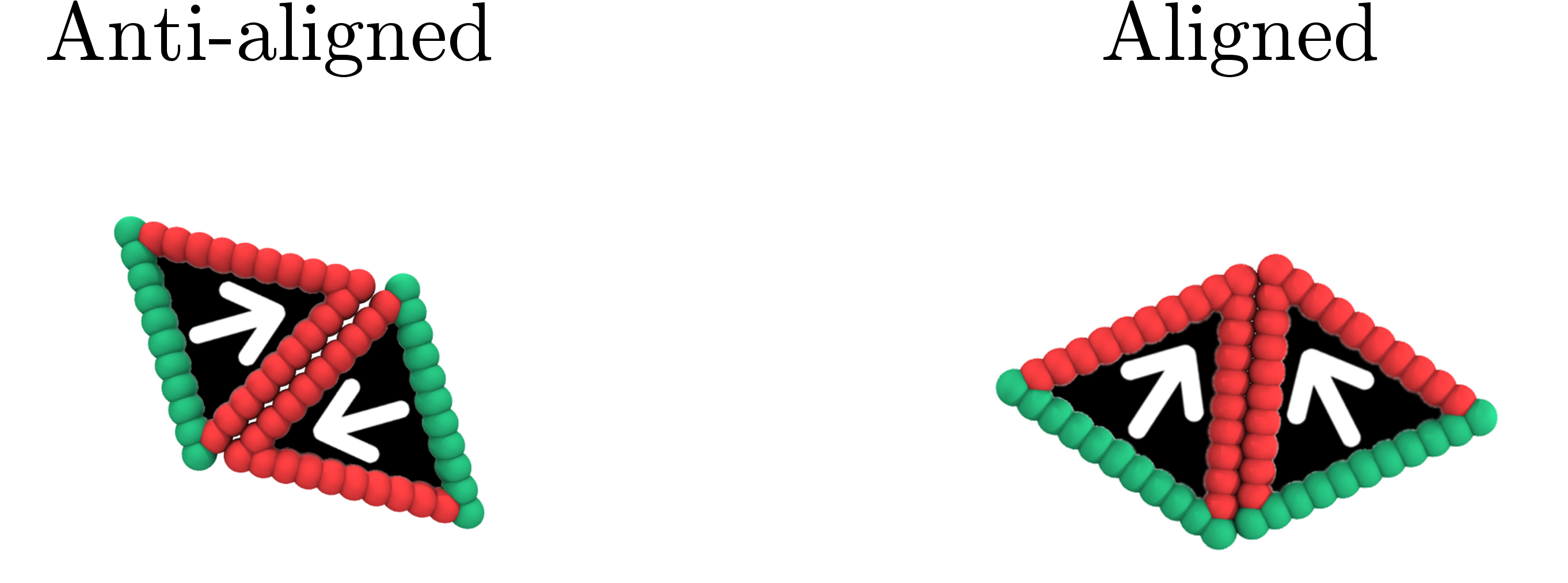}
\caption{(A) Anti-aligned and (B) aligned binding configurations for this choice of colloidal building block.  The introduction of self-propulsion stabilized the aligned configuration and destabilize the anti-aligned configuration.}
\label{fig4}
\end{figure}

It should also be stressed that self-propulsion not only stabilizes the final hexagonal aggregates, but it also biases the formation of specific pair interactions early on in the self-assembly process. As illustrated in Fig.~\ref{fig4}, there are two ways for colloids to bind to each other. In the first configuration, the propelling axes are anti-aligned, whereas in the second configuration the self-propelling axes are partially aligned. In the former case, activity tends to destabilize and separate the pair by creating a shear along the contact edge, in the latter case the bond between the two particles is strengthened and they will move as a pair at a reduced speed whose value depends essentially on the geometry of the particles.  Thus the selection of the final structure begins already at an early stage as the partially aligned configuration is favored and is compatible with the target structure.

One particularly appealing aspect of active colloids is that in many cases the self-propulsion mechanism can be turned on and off by an external light source. \cite{buttinoni284129,cohen068302,palacci20130372,palacci936}.   In what follows, we study how the self-assembly process is affected when allowing the self-propelling force of the particles to change over time. Specifically, we considered a periodic step function where the self-propelling force is turned on for a time $\tau_{\rm on}$ and then turned off for a time $\tau_{\rm off}$ for a total period of $\tau_{0}=\tau_{\rm on}+\tau_{\rm off}$.  For this particular form of the self-propelling force, it is possible to compute exactly the mean square displacement and the swim pressure of an active Brownian particle. These calculations are included in Appendix B.  In the limit for $\tau_{0} \rightarrow 0$, 
\begin{align}
\langle x^2(t)\rangle &\simeq\left[2D+\frac{1}{D_r}\left(\frac{v_0\tau_{\rm on}}{\tau_0}\right)^2\right]t \\
\frac{\Pi^{\rm Swim}}{\rho} & \simeq\frac{\gamma}{D_r} \left(\frac{v_0\tau_{\rm on}}{\tau_0}\right)^2
\end{align}
\noindent where $v_0 = F_a / \gamma$ is the bare swimming speed of the colloid,$\gamma$ is the friction coefficient, and $\rho$ is the colloid density. Both expressions have the functional form of a brownian active particle with a constant propelling force of effective strength equal to  
\begin{equation}
F_{\rm eff}=\frac{\tau_{\rm on} F_a}{\tau_{\rm on}+\tau_{\rm off}}=\frac{\tau_{\rm on} F_a}{\tau_{0}}
\label{eff}
\end{equation}
\noindent which is simply the time averaged force within a single period.  This result is significant as it shows that by tuning the relative on/off time of the self-propelling force it is possible to modulate the effective speed of the colloids. In practical terms, this protocol can be used to modulate the speed of the colloid  which would otherwise be exclusively controlled by the chemical details of the propulsion mechanism.  

\begin{figure}[t!]
\includegraphics[width = 3.0in]{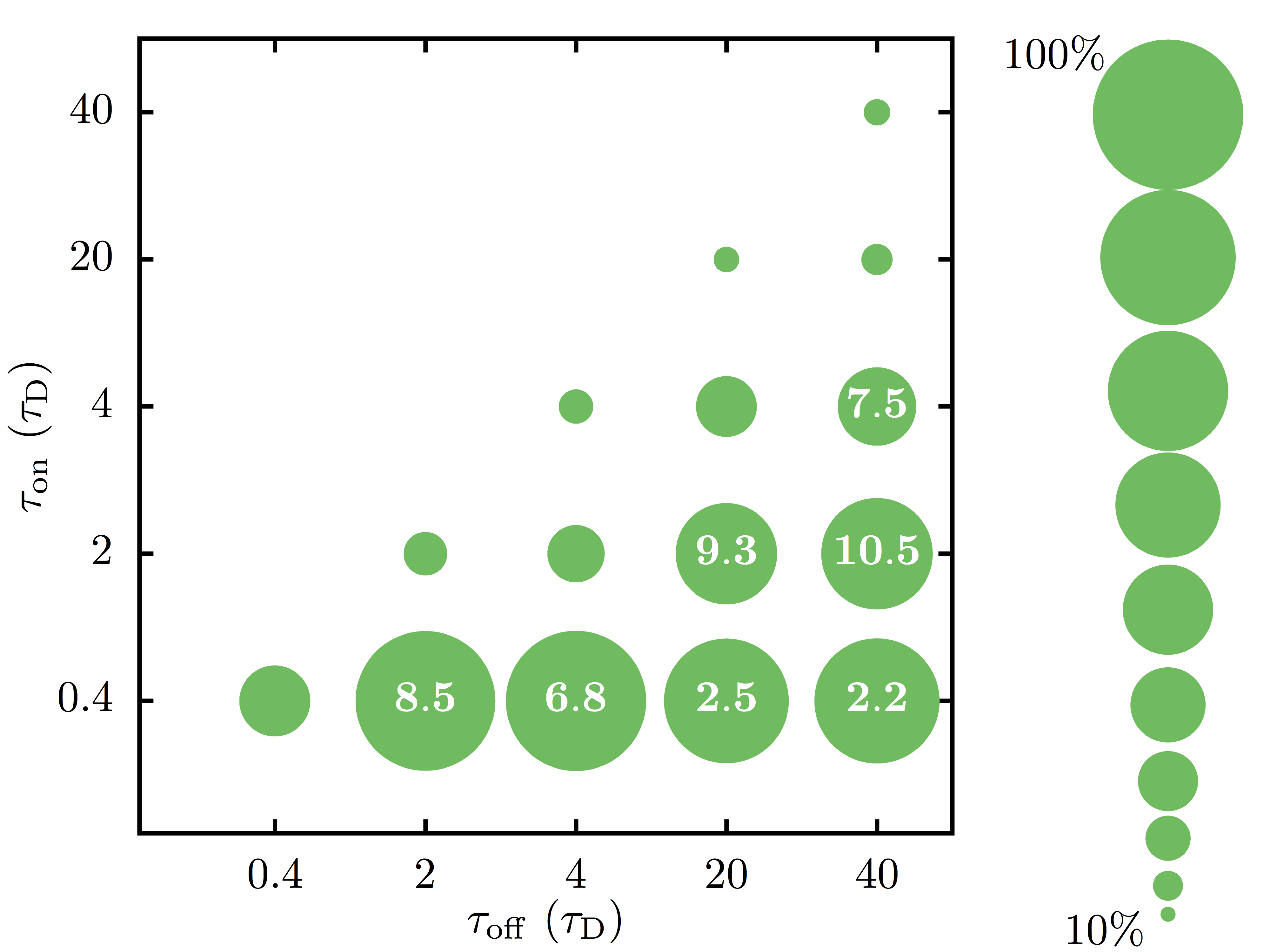}
\caption{Percent yield for colloids with a periodic self-propelling force.  The self-propelling force $F_a \sigma =25 \  k_BT$ is turned on for a time $\tau_{\rm on}$ and then turned off for a time $\tau_{\rm off}$.  The interaction energy is fixed at $\varepsilon \approx 16 \  k_BT$. The white number at the center of the circle indicate the relative increase in speed of the assembly process $\nu_{\rm SA}$.}
\label{fig5}  
\end{figure}

As our example case, we consider a self-propelling force $F_a \sigma =25 \  k_BT$ and interaction energy $\varepsilon \approx 16 \  k_BT$. For this set of parameters the behavior of the system is highly dynamic and the target structures are continually being destroyed by errant free swimming colloids. The yield reaches a maximum value of about $20\%$ when the self-propelling force is constant. When setting $F_a \sigma =0 \   k_BT$, we obtain the best passive yield of about $90\%$. In Figure \ref{fig5}, we plot the percent yield for different pairings of $\tau_{\rm on}$ and $\tau_{\rm off}$. We first discuss the effects of a symmetric periodic step function, i.e. $\tau_{\rm on} = \tau_{\rm off}$.  For large values of $\tau_{\rm on}$ and $\tau_{\rm off}$, which in our simulations corresponds to $\tau_{\rm on}=\tau_{\rm off}=40\,\tau_{\rm D}$, the system gives a yield that is comparable with that obtained for a constant self-propelling force $F_a \sigma = 25 \  k_BT$. At the opposite extreme, for very fast switching times, $\tau_{\rm on}=\tau_{\rm off}=0.4\,\tau_{\rm D}$, we obtain a yield consistent with a constant self-propelling force of half its original value (i.e. $F_a \sigma = 12.5 \  k_BT$). This result is in agreement with our theoretical expectations concerning the effective self-propelling force given in Eq. \eqref{eff}. We find the best yields are obtained when $F_a$ is turned on for $\tau_{\rm on}=0.4\,\tau_{\rm D}$ and then turned off for $\tau_{\rm off}=2\,\tau_{\rm D}$, or greater. This is in good agreement  with our data in Fig.~\ref{fig3}, for which a constant propelling force $F_a \sigma \approx 5 \  k_BT$ gives the best yield for $\varepsilon \approx 16 \  k_BT$.  These results further bolster our argument that the effective force of the colloid can be modulate to its optimal value by tuning $\tau_{\rm on}$ and $\tau_{\rm off}$ for sufficiently small $\tau_0$.  Since $\varepsilon \approx 16 \  k_BT$ is the optimal condition for the passive case, larger values of $\tau_{\rm off}$  effectively decreases the strength of the effective force slowly moving the system towards the passive limit with a roughly constant yield but decreasing $\nu_{\rm SA}$. Finally, we observe that by moving along the diagonals of Fig. \ref{fig5} (points for which $\tau_{\rm on}/\tau_{\rm off}=\rm {const}$) the best yield is obtained for the smallest values of $\tau_0$, (i.e. in the limit where Eq.~\ref{eff} becomes more accurate). In general, increasing $\tau_0$, while keeping the ratio $\tau_{\rm on}/\tau_{\rm off}$ fixed, decrease the overall yield, but may lead to faster rates of self-assembly.

In summary, we have discovered that a judicious use of self-propulsion can greatly benefit the colloidal self-assembly of a certain class of target micro-structures. In simple terms, we show that fast and sticky colloids can successfully self-assemble more than one order of magnitude faster than their passive counterparts without sacrificing the overall yield.  We also demonstrate how tuning the relative on/off time of the self-propelling force (i.e. by using quick bursts of activity rather than a constant force) it is possible to modulate the effective speed of the colloids allowing for further optimization of the self-assembly process. This result also suggests that quickly toggling the self-propelling force is a simple method to control the microscopic speed of the colloids and it is not always necessary to tinker with the chemical details of the propulsion mechanism. 

Although we study this problem through the lens of a simple minimal model, we believe that the general approach and methodology discussed here can be successfully applied to other colloidal systems with different building blocks as long as the necessary criteria are satisfied.  We don't believe that the vectorial constraint discussed above needs to be strictly satisfied, and aggregates whose propelling axes sum up to a sufficiently small value would still see an improvement in self-assembly. For instance small clusters of Janus particles in three dimensions, should be amenable to this approach. More work in this direction is needed.  

{Finally,  we should stress that at low densities, hydrodynamic interactions can lead to long range forces mediated by the surrounding  fluid. In this work, we haven't explicitly considered this effect, however, we expect it to lead mostly to quantitative and not qualitative changes in the phase behavior as long as hydrodynamic intractions do not break the stable aggregates. Recent work \cite{zottl118101} indicates that hydrodynamics has a crucial role on the rotational dynamics of spheres inside living crystals, however, in our case the anisotropic nature of the  direct interparticle interactions and their excluded volumes forbids them to rotate once assembled, which negates this effect. Additional work is needed to understand how hydrodynamics interactions may affect the self-assembly pathway in active systems.}

\begin{acknowledgments}
We thank Clarion Tung, Joseph Harder and Chantal Valeriani for insightful discussions and helpful comments. AC acknowledges financial supported from the National Science Foundation under Grant No. DMR-1408259. SAM acknowledges financial support from the National Science Foundation Graduate Research Fellowship grant number DGE-11-44155. This work used the Extreme Science and Engineering Discovery Environment (XSEDE), which is supported by National Science Foundation grant number ACI-1053575.
\end{acknowledgments}

\appendix

\section{Additional simuation details}

For computational efficiency, each colloid is discretized into $N_T = 30$ equidistant, partially overlapping, and rigidly connected spherical subunits of diameter $\sigma_T=0.25 \sigma$, where $\sigma$ is the unit length in our simulations.  The edge length of the colloid is fixed at $2 \sigma$.  A suitably large number of spheres was chosen to accurately reproduce the shape of the colloid and to make the surface interaction sufficiently smooth. All particle interactions in the systems are given by a Lennard-Jones potential
\begin{equation}
V(r)=4 \varepsilon_{ij} \left[ \left( \frac{\sigma_{T}}{r} \right)^{12}- \left( \frac{\sigma_{T}}{r} \right)^{6} \right]
\end{equation}
The spherical subunits that make up the colloid come in two different varieties: type A and  type B, and the $i$ and $j$ indices refer to them. Type A particles are responsible for the attractive interaction while type B only account for volume exclusion. The distinguishing characteristic between these two particle types is the potential cutoff distance. The cutoff distance between type A particles is set to $r_c = 2.5 \sigma_T$. Type A particles are in essence Lennard-Jones particles and exhibit a weak, short range attraction, which is modulated by the interaction energy $\varepsilon_{AA}$. The cut off distance for the remaining pair interactions (type A-type B and type B-type B) is set to $r_c = 2^{1/6} \sigma_T$ making these interactions purely repulsive in nature.  The interaction energy for these volume excluding pair interactions is fixed at $\varepsilon_{BB}=\varepsilon_{AB}=1$. In both cases the potential is shifted after truncation so that $V(r_c)=0$.

Each colloid is confined to move in two dimensions, has mass $m$, and undergoes Langevin dynamics at a constant temperature $T$. At each time-step, both the total force and torque on each rigid body is computed as the sum of the forces and torques on its constituent particles.  The coordinates, velocities, and orientations of the particles in each body are then updated so that the body moves and rotates as a single entity.  Self-propulsion is introduced through a directional force of constant magnitude $F_a$ and is directed along a predefined orientation vector $\boldsymbol{n}$ which passes through the center of mass (COM) of each particle and is perpendicular to the purely repulsive face of the colloid as illustrated in Fig. 1A of the main text.  The equations of motion of an individual colloid are given by the coupled Langevin equations

\begin{align}
m \ddot{ \boldsymbol{r}_i} & =-\gamma \dot{ \boldsymbol{r}_i}+F_a{\boldsymbol n}_i-{\boldsymbol \nabla}_iV(r_{ij})+\sqrt{2\gamma^2D} \,{\boldsymbol \xi}_i \\
 \boldsymbol{\dot{n}}_i& = -\frac{D_r}{k_{\rm B}T}  \boldsymbol{{\nabla}_{\boldsymbol{n}_i}} V(r_{ij})+\sqrt{2D_r}\,{\boldsymbol \xi}_{r_i}\times  \boldsymbol{n}_i
 \end{align}
\noindent where $\gamma$ is the translational friction and $V$ the total interparticle potential acting on the particle. The translational and rotational diffusion constants are given by $D$ and $D_r$, respectively. The typical solvent induced Gaussian white noise terms for both the translational and rotational motion are characterized by $\langle \xi_i (t)\rangle = 0$ and $\langle \xi_i(t) \cdot \xi_j(t') \rangle = \delta_{ij}\delta(t-t')$, $\langle \xi_{r}(t)\rangle = 0$ and $\langle \xi_{r}(t) \cdot \xi_{r}(t') \rangle =\delta(t-t')$. The translational diffusion constant $D$ is related to the temperature $T$ via the Stokes-Einstein relation $D=k_BT/\gamma$.  All quantities in this investigation are given in reduced Lennard-Jones units. The friction coefficient $\gamma$ was chosen such that the translational and rotational motion of the colloid is overdamped. 

\section{Derivation of MSD and Swim Pressure for oscillating Swimming Speed}

For simplicity, we consider an active Brownian particle in one dimension
\begin{align}
\dot{x}(t) & = v_a(t){\cos[\theta(t)]} +\sqrt{2D}\,\,{\bf\xi}(t) \\ 
\dot{\theta}(t) & =\sqrt{2D_r}\,\,{\bf\xi}_r(t)
\end{align}
where the $\xi$ and $\xi_r$ correspond to the gaussian distributed translational and rotational random noises with $\langle \xi (t)\rangle = 0$, $\langle \xi(t) \cdot \xi(t') \rangle = \delta(t-t')$ and $\langle \xi_{r}(t)\rangle = 0$, $\langle \xi_{r}(t) \cdot \xi_{r}(t') \rangle =\delta(t-t')$, respectively. The propelling velocity $v_a(t)$ has the functional form of a asymmetric periodic step function $v_a(t)=v_0 \Xi[t]$ with
\begin{equation}
\Xi(t)=
\begin{cases}
1 \,\,\,\,\, {\rm if }\,\,\,\,\, n(\tau_{\rm on}+\tau_{\rm off})\leq t\leq  n(\tau_{\rm on}+\tau_{\rm off}) +\tau_{\rm on}\\
0 \,\,\,\,\, {\rm Otherwise }
\end{cases}
\end{equation}
with period $\tau_0=\tau_{\rm on}+\tau_{\rm off}$, $n=0,1,2,3, ....,t/\tau_0$, and $\theta(t)$  is the angle between the $x$ axis and the propelling axis of the particle. Clearly, $\tau_{\rm on}$ is the time the propelling velocity is active and $\tau_{\rm off}$ is the time off, and $v_0$ is the bare propulsion speed of the particle.  To draw comparison with our simulations of the triangular colloids $v_o =F_a/\gamma$.  We can now directly compute the mean square displacement of the particle

\begin{align}
\nonumber
 \langle x^2(t)\rangle & = v_0^2\int_0^t \int_0^t \Xi(t)\Xi(t')  \\  &  \times \langle \cos[\theta(t)]\cos[\theta(t')]\rangle \,dtdt' +2Dt
\end{align}

\noindent where $\langle \cos[\theta(t)]\cos[\theta(t')]\rangle_{t > t'}=\frac{1}{2} e^{-D_r(t-t')}$. It follows that

\begin{widetext}
\begin{equation}
\langle x^2(t)\rangle=2Dt+\frac{v_0^2}{2} \left [\int_0^tds\, e^{-D_rs} \Xi(s) \int_0^s dp\,  e^{D_rp} \Xi(p) +
\int_0^tds \,e^{D_rs} \Xi(s) \int_s^t dp \,e^{-D_rp} \Xi(p) \right ]
\label{doubleI}
\end{equation}
\end{widetext}

\onecolumngrid
\noindent We begin by evaluating the first double integral on the right hand side of the equation above. The integral over $p$, which we define as $G_1(s)$ can be rewritten as a geometric series

\begin{widetext}
\begin{equation}
G_1(s)=\int_0^s dp\,  e^{D_rp} \Xi(p)=\frac{(e^{D_r\tau_{\rm on}}-1)}{D_r}\sum_{n=0}^{\frac{s}{\tau_0}-1} e^{nD_r\tau_0}=
\frac{(e^{D_r\tau_{\rm on}}-1)}{D_r}\frac{(1-e^{D_rs})}{(1-e^{D_r\tau_0})}
\end{equation}
\end{widetext}

\onecolumngrid
\noindent By substituting this expression for $G_1(s)$ we are able to integrate the complete double integral which we define as $Q_1(t)$

\begin{widetext}
\begin{align}
Q_1(t) & =\int_0^tds\,e^{-D_rs}\Xi(s)G_1(s)=\frac{(e^{D_r\tau_{\rm on}}-1)}{D_r(1-e^{D_r\tau_0})} \int_0^tds\,(e^{-D_rs}-1)\Xi(s) \\
\nonumber
&=\frac{(1-e^{D_r\tau_{\rm on}})}{D_r(1-e^{D_r\tau_0})}\left [
 \frac{\tau_{\rm on}}{\tau_0}\,\,t +\frac{(e^{-D_r\tau_{\rm on}}-1)}{D_r} \sum_{n=0}^{\frac{t}{\tau_0}-1} e^{-nD_r\tau_0} 
 \right ]\, \\
\nonumber
& =\frac{(1-e^{D_r\tau_{\rm on}})}{D_r(1-e^{D_r\tau_0})}\left [
 \frac{\tau_{\rm on}}{\tau_0}\,\,t +\frac{(e^{-D_r\tau_{\rm on}}-1)(1-e^{-D_rt})}{D_r(1-e^{-D_r\tau_0})}\right ]
\end{align}
\end{widetext}
\onecolumngrid
\noindent In a similar fashion we can compute the second double integral on the right hand side of Eq.~\ref{doubleI} which we define as $Q_2(t)$

\begin{widetext}
\begin{equation}
Q_2(t)= \frac{(1-e^{-D_r\tau_{\rm on}})}{D_r(1-e^{-D_r\tau_0})}
\left [ \frac{\tau_{\rm on}}{\tau_0}\,\,t   + \frac{(e^{D_r\tau_{\rm on}}-1)}{D_r(1-e^{D_r\tau_0})}(1-e^{-D_rt}) \right ]
\end{equation}
\end{widetext}

\noindent The expression for the mean square displacement can now be rewritten as 

\begin{equation}
\langle x^2(t)\rangle=2Dt+\frac{v_0^2}{2}
\left[Q_1(t)+Q_2(t)
\right]
\end{equation}

\noindent In the diffusive limit $(t \rightarrow \infty)$ with a short total period ($\tau_0 \rightarrow 0$), we find apart from a constant

\begin{equation}
\langle x^2(t)\rangle\simeq\left[2D+\frac{1}{D_r}\left(\frac{v_0\tau_{\rm on}}{\tau_0}\right)^2\right]t \,\,. 
\label{diff}
\end{equation}

\noindent In the ballistic limit $(t \rightarrow 0)$ with a short total period ($\tau_0 \rightarrow 0$), we find 

\begin{equation}
\langle x^2(t)\rangle\simeq \left(\frac{v_0\tau_{\rm on}}{\tau_0}\right)^2t^2 \,\,. 
\end{equation} 

\noindent In both cases, when $\tau_{\rm on}=\tau_0$ one recovers the well known results for a constant self-propelling force.  It is now possible to obtain the swim pressure from the stress $\Pi=-{\rm Tr}\,\, \sigma^{\rm swim}/3$ by simply computing $\sigma^{\rm swim}=- \langle x F^{\rm swim}\rangle$.  In the diffusive limit the swim pressure is 

\begin{equation}
\frac{\Pi^{\rm Swim}}{\rho}=\frac{\gamma}{D_r} \left(\frac{v_0\tau_{\rm on}}{\tau_0}\right)^2
\end{equation}

\noindent where $\rho$ is the colloid density.

\bibliographystyle{apsrev4-1}
\bibliography{hex.bib}

\end{document}